# Multicarrier Spectral Shaping for Non-White Interference Channels: Application to L-band Aviation Channels

Hosseinali Jamal, *Member IEEE*, David W. Matolak, *Senior Member IEEE*

*Abstract*—In this paper, we investigate an algorithm to attain additive white Gaussian noise (AWGN) performance in spectrally non-white channels, using multicarrier communications, i.e., orthogonal frequency division multiplexing (OFDM) and filterbank multicarrier (FBMC). The non-white channel can be from non-white noise, or more commonly, interference. Our algorithm uses a simple optimization method to find usable subcarriers and assigns different power levels to the subcarriers to attain the equivalent AWGN channel bit error ratio (BER). Subcarriers that experience very high interference are assigned as null subcarriers. After describing our analysis, we show results for two non-white interference signal examples: the Gaussian pulse shaped distance measuring equipment (DME) pulses, and a classical rectangular-pulse interference signal. The DME example is pertinent for currently proposed aviation communication systems, where new multicarrier techniques, e.g., the L-band digital aviation communication systems (LDACS) have been designed as an inlay approach between the high-power DME channels in the L-band. Our results show how using this adaptive technique can improve performance and spectral efficiency, whereas fixed bandwidth schemes such as LDACS could suffer significant performance degradation. These results show the utility of this idea for future adaptive and cognitive radio applications for aviation or other non-white channels.

# I. INTRODUCTION

The use of adaptive radio techniques may have its roots in the 1960s with adaptive equalization [1]. Adaptive transmit power control was used by AT&T in the 1980s, and cellular radio introduced numerous additional adaptive applications in subsequent years. Use of modern communication technologies in aviation lags that in commercial applications such as cellular and wireless local area networks (WLANs). Most parts of dedicated aviation spectrum are occupied by traditional analog communication technologies such as double-sideband amplitude modulation (DSB-AM) with no carrier suppression, currently used for voice communication in the VHF band. Similarly, the pulsed system distance measuring equipment (DME) is a more than 50-year-old technology used for aircraft slant-range measurement.

Digital communication-based systems were recently introduced for aeronautical communications, such as VHF digital link modes 2 and 3 (VDL2/VDL3), and L-band digital aeronautical communication systems (LDACS). Digital communications offer the potential to improve the capacity and capability of the communication system, thereby improving air traffic management efficiency and safety. Adaptive physical link technologies are of interest for aeronautical systems to further improve performance.

The number of aviation communication system users is increasing, especially for unmanned aerial vehicle (UAV) systems. Therefore, demands for communication systems with greater reliability and higher communication capacity are increasing. For civil aviation communication systems the largest spectrum allocations are in a few bands: VHF (108-137 MHz) for transmitting real-time data (mostly voice) for air-traffic control (ATC), air-traffic management (ATM), and air traffic services, and the frequency channels in L-band (960-1215 MHz), currently primarily for aviation radio-navigation and surveillance purposes. Bands from 5.03-5.091 and 5.091-5.15 GHz are also being considered for future UAV and airport surface use. Statistically speaking, the spectrum utilization in these bands for a specific geographical location is very low, especially for the L-band channels. Therefore, new data services for aircraft passenger communications (APC) and airline operation communications (AOC) are expected to be a major motivation for broadband air-to-ground (AG) communication systems [2] in this band.

In [3] we provided a review of currently used communication and transponder systems in L-band. One system that is widely used throughout the L-band is DME. DME signals are high power pulsed signals used for radio-navigation. The DME signal pulses are Gaussian in shape, and hence yield non-white (Gaussian shaped) spectra. As we will show, the distribution of DME channels in frequency leaves large segments of the spectrum vacant; we term these segments spectral gaps.

The idea of using the spectral gaps between DME channels for broadband communication purposes was first proposed by EUROCONTROL, using an orthogonal frequency division multiplexing (OFDM) based system, LDACS-1 [4] (we neglect the LDACS-2 technology, which is no longer being studied, and abbreviate LDACS-1 as LDACS). In some extreme and relatively high-power situations, DME can significantly degrade the LDACS performance; hence, there have been studies i.e., [5]-[8], proposing techniques such as pulse clipping/blanking to deal with DME interference. According to the results in these studies, simple techniques such as pulse blanking cannot effectively deal with DME (1 to 2 dB improvement in SNR), and even with complex techniques in [7] and [8], there are remains several dB degradation compared to the additive white Gaussian noise (AWGN) channel results. In this paper, we study an adaptive algorithm that enables us to attain AWGN channel results without using any interference mitigation technique at the receiver.

In [3] we designed a new communication system based on filterbank multicarrier (FBMC) following LDACS requirements, and via analysis and simulations, we showed that the FBMC technique has some notable advantages over LDACS, primarily in terms of throughput and robustness to DME interference. In [9] we expanded the FBMC design via a spectrally shaped FBMC (SS-FBMC) system. In that initial SS-FBMC work we devised an algorithm to equalize the overall bit error ratio (BER) performance across subcarriers in the presence of DME transmission. In this paper, we modify our previous algorithm to attain theoretical AWGN channel (non-interference) performance. This does not mean we remove the interference at receiver, but instead we allocate the subcarriers, which can attain AWGN performance.

In recent years, some standardization activities, such as IEEE 802.22, have contributed to communication systems based on cognitive radio (CR) for wireless regional area networks (WRAN) and terrestrial networks [10]. The idea of CR in aviation systems has been described in [11] for the VHF bands. Later in [2], the authors expanded this idea with more sophisticated algorithms and additional results for the same band. It is known that FBMC is suitable for CR: due to sharp subcarrier filtering it can provide fine frequency resolution to detect subcarrier power levels for spectrum sensing and power allocation purposes [12]-[15]. In this paper, we generalize the idea of SS-FBMC and modify our algorithm for SS-FBMC that can be applied to any non-white interference channels. To apply our algorithm in practice would require the non-white spectrum information at the transmitter, which can be provided by using spectrum-sensing techniques as in CR. References [16] and [17] are two recently published papers related to the LDACS spectrum sensing at L-band. For simulations of this technique, we chose two non-white channel spectra examples: DME signals and classic rectangular pulses. Via our analysis and simulation results, we show the advantage of spectrally shaped (SS) schemes as a practical approach for non-white channels such as the DME interference channel; this is in comparison to LDACS, which incurs significant performance degradation.

The rest of this paper is organized as follows: in Section II, we describe the spectral shaping algorithm for non-white interference channels. In Section III we describe the two interference signal models we use as non-white interference examples for analyzing the proposed SS-FBMC technique: the DME, and rectangular pulses with the well-known $[\sin(\pi fT)/\pi fT\,]^2$ spectrum. In Section IV we briefly review L-band communication system candidates LDACS and the SS schemes. Section V presents power spectral density (PSD), and BER simulation results for the SS-FBMC/OFDM and LDACS systems for comparison, and Section VI concludes the paper.

## II. SPECTRAL SHAPING (SS) ALGORITHM

In [9] we proposed an optimization method to transmit a low power FBMC signal in the spectral gaps of two adjacent DME channels to equalize BER and remove any error floors due to DME interference. As later shown, the DME PSD is largest at the DME center frequencies, and "gaps" are where the spectral density is low. In this paper, we generalize the idea of [9] in order to use it in any non-white spectrum channel, with the aim of attaining AWGN channel theoretical BER results.

As the main metric in this algorithm and first step, we used Shannon's channel capacity theorem[1] [18]. By this technique, we can determine subcarriers suitable for transmission versus those not. This means, some subcarriers incurring the largest interference (nearest the DME channel centers) will be nulled. In the second step, considering the interference on each remaining active (data) subcarrier, we distribute the total power among the data subcarriers in a way to obtain theoretical AWGN performance on those subcarriers. After applying this algorithm some subcarriers with larger interference density might have larger power levels than others, therefore, the spectrum is shaped and is non-flat, compared to the traditional multicarrier systems such as LDACS, which have a mainly flat spectrum.

---

[1] We note that Shannon's capacity formula pertains to white noise channels, and our interference densities are non-white. For a sufficiently small subcarrier bandwidth, on a per-subcarrier basis the interference density approaches a white form.

To help illustrate the spectral shaping algorithm, we plot an example half bandwidth PSD of two identical adjacent channels, generated from rectangular pulses in Figure 1. As mentioned, the idea is using the spectrum gap between these channels in order to transmit our SS waveform with $N$ subcarriers. Thus, the center frequencies of these two channels are aligned with the subcarrier indices 1 and $N$. We assume an arbitrary example subcarrier bandwidth, $\Delta f$ Hz, which is of course related to the bandwidth of the rectangular pulse signals. The shaded area (red) indicates the channel interference $I_n$ on subcarrier $n$'s band; this is calculated by integrating the PSD over this subcarrier $n$ bandwidth.

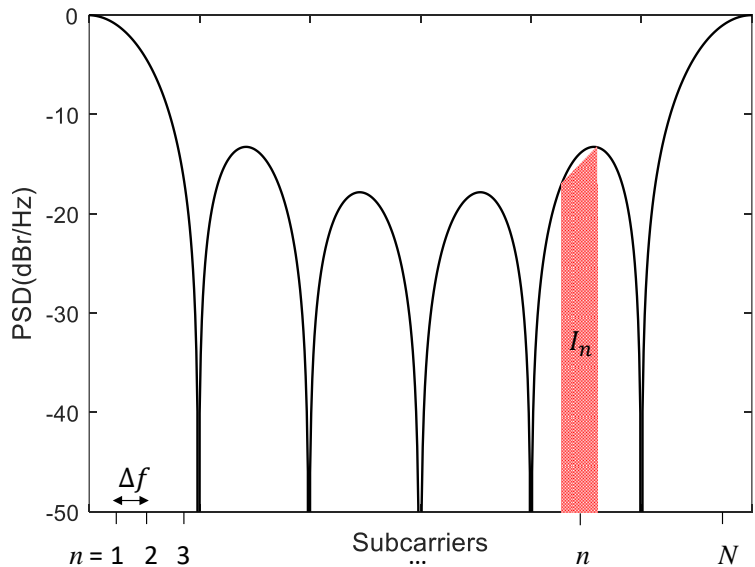

Fig. 1. Example $sinc^2$ PSD vs. subcarrier index.

For subcarrier $n$, based on Shannon's capacity theorem assuming an additive white Gaussian noise (AWGN) channel, the capacity can be approximated as

$$C_n \cong \Delta f \log_2\left(1 + \frac{P_n}{\gamma_n^2}\right) \quad n = 1, \dots, N, \tag{1}$$

where $P_n$ is the power allocated on subcarrier $n$, and $\gamma_n^2 = I_n + \Delta f \sigma_{AWGN}^2$ is the interference power plus noise power at subcarrier $n$. Thus, $P_n/\gamma_n^2$ is the signal to interference plus noise ratio (SINR) at subcarrier $n$. As noted, we can estimate $I_n$ as follows,

$$I_n = \int_{f_n - \Delta f/2}^{f_n + \Delta f/2} |S(f)|^2 \, df, \quad n = 1, \dots, N, \tag{2}$$

where $|S(f)|^2$ is the PSD of the interfering signal, and $f_n$ is the center frequency of subcarrier $n$.

After estimation of the interference density, our algorithm is formulated in the following optimization problem, with $P_T$ the total available transmit power:

$$\underset{\alpha_n, P_n}{Max} : \sum_{n=1}^{N} \alpha_n C_n, \tag{3}$$

*subject to*:

    1: $P_n \geq 0$,

    2: $\alpha_n \in \{0,1\}$,

    3: $\alpha_n = 0, \ n \in \{1, 2, \dots, \frac{N}{4} + 1, \frac{N}{2} + 1, \frac{3N}{4} + 1, \dots, N\}$,

    4: $\sum_{n=1}^{N} P_n = P_T$,

    5: $\forall n | \alpha_n = 1: I_n < K\Delta f \sigma_{AWGN}^2$.

This optimization problem has five constraints, 1: power allocated to each subcarrier is positive or zero; 2: parameter $\alpha_n$ is defined for subcarrier allocation, which is either 0 or 1; 3: for practical implementations, we assume zero or null DC subcarrier, also following LDACS we assume the effective bandwidth never exceeds half the total DME bandwidth (0.5 MHz in our simulations), thus the first and last quarter of the subcarriers are always set zero. Note this ¼ is an arbitrary number, and choosing half of the bandwidth satisfies the maximum LDACS usable bandwidth (0.5 MHz); 4: total communication system transmit power is set $P_T$; and, 5: for those active subcarriers $n$ where $\alpha_n = 1$, in order to satisfy the AWGN channel performance, the interference should be less than a threshold that is related to the noise power within each subcarrier band. Here we included the factor $K$ in the threshold as a design parameter (i.e. for $K \ll 1$, $\gamma_n^2 \cong \Delta f \sigma_{AWGN}^2$ which resembles an AWGN channel). This factor will be analyzed in the simulation results section. We note that this optimization problem is convex and has a unique solution. Hence in order to check our water-filling algorithm; we also used CVX (a MATLAB® package for specifying and solving convex problems) [19]. The CVX code is also provided in the Appendix for reference. We assured that the CVX results are the same as those found by our proposed solution algorithm.

These kind of optimization problems have a well-known solution based on the water-filling algorithm [20]. In the Appendix, we provide a computationally efficient solution to solve this problem. This algorithm provides subcarrier allocation, which is required to find the set of "active subcarriers", i.e. $n \in \{1, 2, \dots N\}$ such that $\alpha_n = 1$. Subcarriers with $\alpha_n = 1$ are chosen as active subcarriers and the others are made null subcarriers. This algorithm enables us to adaptively find the guard band and to shape the transmitting spectrum of our waveform according to the non-white channel spectrum. This "interference-aware" operation adapts to the interference, hence improving performance and maximizing the reliability, which is critical for aeronautical communication systems.

## III. EXAMPLE NON-WHITE SPECTRA FOR SS ALGORITHM ANALYSIS

In this section, first we describe the non-white spectrum of the DME signal since it is our primary practical interference signal example. Then we describe a second example non-white spectrum signal based on rectangular pulses, also used for analysis.

### A. DME Pulses

DME is a traditional transponder-based navigation technology over 50 years old, used to measure slant range distance to civil aircraft by calculating the timing of propagation delays. DME ground stations are distributed geographically at known locations, like cellular base stations. Each single DME channel occupies 1 MHz bandwidth. Based on United States frequency allocations, DME frequency channels are allocated at 1 MHz frequency increments throughout the 960 to 1150 MHz band. There are about 152 DME channels being used at more than 1100 DME ground stations in the U.S.

The DME signals are a sequence of Gaussian pulse pairs with the pulse start times modeled as a Poisson process. DME signals are high power signals with peak power of 1000 W for ground station transmitters and 300 W for aircraft. The pulse pair transmission rate of DME systems varies: the maximum rate from ground station to the aircraft is 2700 pulse pairs per second (ppps), and for the aircraft to ground station 150 ppps. Each DME pulse pair can be generated as in (4), where the two pulses are separated by $\Delta t$ and the constant $\beta$ determines the pulse width. Note that the start time of each pulse pair is modeled following a random Poisson process, where in (4) assumed zero

$$S_{DME}(t) = \sqrt{P_{peak}}e^{-\frac{\beta\left(t-\frac{\Delta t}{2}\right)^2}{2}} + \sqrt{P_{peak}}e^{\frac{-\beta\left(t-\frac{3\Delta t}{2}\right)^2}{2}}, 0 \leq t \leq 2\Delta t. \tag{4}$$

According to DME specifications $\beta = 4.5 \times 10^{11} \, s^{-2}$, $\Delta t = 12$, or $36 \, \mu s$ and $P_{peak}$ is DME peak power. We calculate the following integral numerically in order to find the mean power of a single pulse pair, which is related to the peak power

$$P_{S_{DME}(t)} = \frac{1}{2\Delta t} \int_0^{2\Delta t} S_{DME}^2(t) dt$$

$$= \frac{\sqrt{\pi} P_{peak} e^{-\frac{\Delta t^2 \beta}{4}} \left[ e^{\frac{\Delta t^2 \beta}{4}} \left( \text{erf}\left(\frac{3\Delta t \sqrt{\beta}}{2}\right) + \text{erf}\left(\frac{\Delta t \sqrt{\beta}}{2}\right) \right) + 2 \, \text{erf}(\Delta t \sqrt{\beta}) \right]}{2 \Delta t \sqrt{\beta}}$$

$$\cong 0.22018 P_{peak}, \quad (5)$$

Then, the mean power of the sequence of DME pulses with rate 2700 ppps is approximately $0.22018 P_{peak} \times 2700 \times 2\Delta t$. After taking the Fourier transform of the DME pulse pair in (4) we have the following result

$$S_{DME}(f) = \sqrt{\frac{8\pi P_{peak}}{\beta}} e^{\left(-\frac{2\pi^2 f^2}{\beta}\right)} e^{(j\pi f \Delta t)} \cos(\pi f \Delta t). \quad (6)$$

Similar to frequency reuse in cellular networks, adjacent DME cells have different channel frequencies, and only one DME channel is allocated for a large coverage area of up to 240 km radius for high altitude aircraft. Except the pulse pair start times, DME signals have deterministic characteristics, and information such as center frequency and power can be detected by spectrum sensing if not already tabulated (and known) by L-band transceivers.

B. *Rectangular Pulses*

As another example interfering signal, we analyze rectangular pulses. In our simulations, in order to generate the sequence of pulses, we assumed the same signal timing process as DME, i.e., the start time is based on a Poisson process. The time domain and Fourier transform for a *single* rectangular pulse are given in (7) and (8). The mean power of a single rectangular pulse is $P_{peak}$, and the power of the sequence of pulses is $P_{peak} \times T \times ppps$. Thus, in order to have similar power as DME, we should have $T = 0.22018 \times 2\Delta t \cong 5 \ \mu s$, which we also chose in our simulations.

$$S_{rect}(t) = \Pi(t) = \begin{cases} \sqrt{P_{peak}}, & 0 \leq t \leq T \\ 0, & t > T \end{cases} \quad (7)$$

$$S_{rect}(f) = \sqrt{P_{peak}} T \ \text{sinc}(Tf). \quad (8)$$

Analytical and simulated half channel PSD results for two identical adjacent channels of these two waveform types are plotted in Figure 2. Thus, we have a plot over 1 MHz bandwidth, where adjacent channels have equal power. As we will explain further regarding simulation results, these PSD power levels are based on a free-space path-loss model and an aviation channel link budget calculation for some given parameters, and a 100 km distance between DME transmitter on the ground and the SS-FBMC/OFDM receiver aircraft. As can be seen, both signals have non-white spectra.

According to the spectral shapes of these interfering signals, the main contribution of this paper is the algorithm to solve (3), hence transmitting a desired SS-FBMC/OFDM signal within the spectral gaps of these signals, in order to attain AWGN channel BER performance. We note that this technique might also be employed in channels without interference or "primary user" signals with colored noise, but using it in the aeronautical case is very practical, as the channel is often dominated by a strong line of sight component, quasi-AWGN, and the DME spectrum is deterministic.

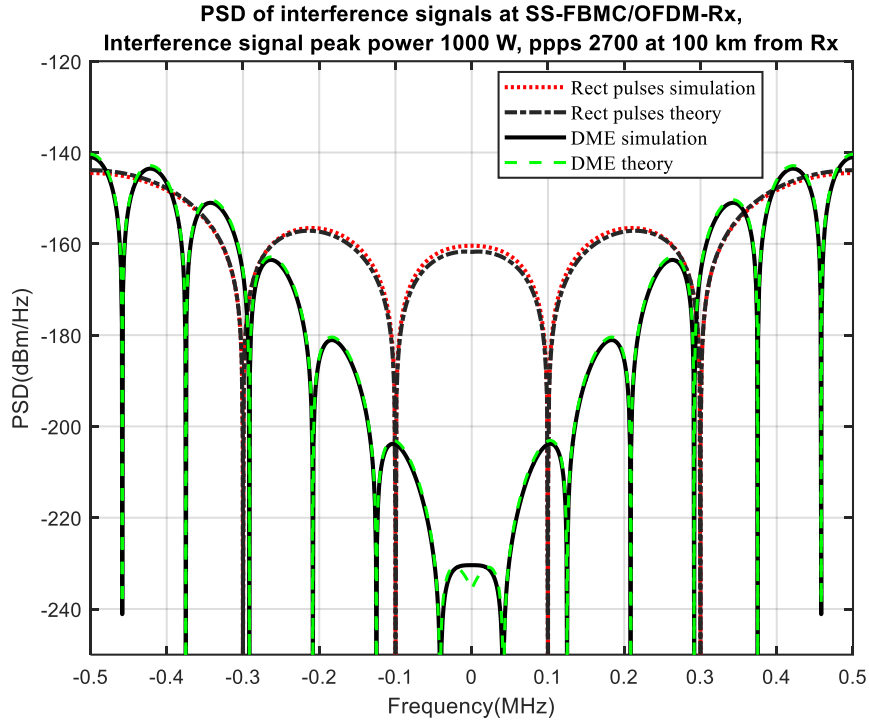

Fig. 2. DME and rectangular pulse PSDs with equal mean power.

## IV. L-BAND AVIATION MULTICARRIER COMMUNICATION SYSTEMS

In this section we briefly describe the EUROCONTROL proposed waveform LDACS CP-OFDM based multicarrier communication system. Then we review our proposed SS-FBMC multicarrier [9] system.

*A. LDACS*

LDACS is a CP-OFDM based communication system like IEEE 802.16 for air-to-ground communications. It has a cellular point-to-multipoint system structure with a star-topology. It uses frequency division duplexing supporting transmission in the forward link (FL, the ground to air channels), and reverse link (RL, the air to ground channels) [4].

The idea of LDACS was first proposed by EUROCONTROL in order to use the DME channels' "spectrum holes" for broadband communication purposes. The bandwidth of LDACS is about 0.5 MHz and fixed with a flat spectrum, thus it is not designed for non-white channels such as DME, and as will be shown, this results in BER floors. For further information about LDACS, refer to [4], [21].

*B. SS-FBMC*

In [3] we proposed an FBMC based communication system for AG communication systems as an alternative to LDACS. We showed that due to the sharper subcarrier filtering, FBMC has several advantages over LDACS, specifically better spectral efficiency (higher throughput), lower out-of-band power emission (lower interference to adjacent channels due to filtering), and more robustness to DME signals. The core system model that we used is based on the staggered multitone (SMT) modulation technique [22]. In [5] we studied the idea of SS-FBMC for L-band NextGen communication systems, providing basic results.

The main idea of SS-FBMC is to shape the spectrum by applying different power levels on subcarriers with weight factors. The weight factors are calculated from (3), and these are the relative power levels on each active subcarrier. The transmission system of SS-FBMC is shown in Figure 3. In this block diagram *N* parallel complex data streams are mapped through OQAM modulation and then passed through subcarrier filters. These signals are then multiplied by a set of weight factors [$A_0$, $A_1$, …, $A_{N-1}$]. Similar diagrams can be drawn for OFDM based spectrally shaped system (SS-OFDM).

For these SS schemes, we can select any total number of subcarriers (size of FFT), thus subcarrier spacing can change depending on the total number of subcarriers. Also, unlike LDACS, the number of guard and used subcarriers, which together define effective bandwidth, is no longer fixed: according to (3) it can be different, depending on interference conditions.

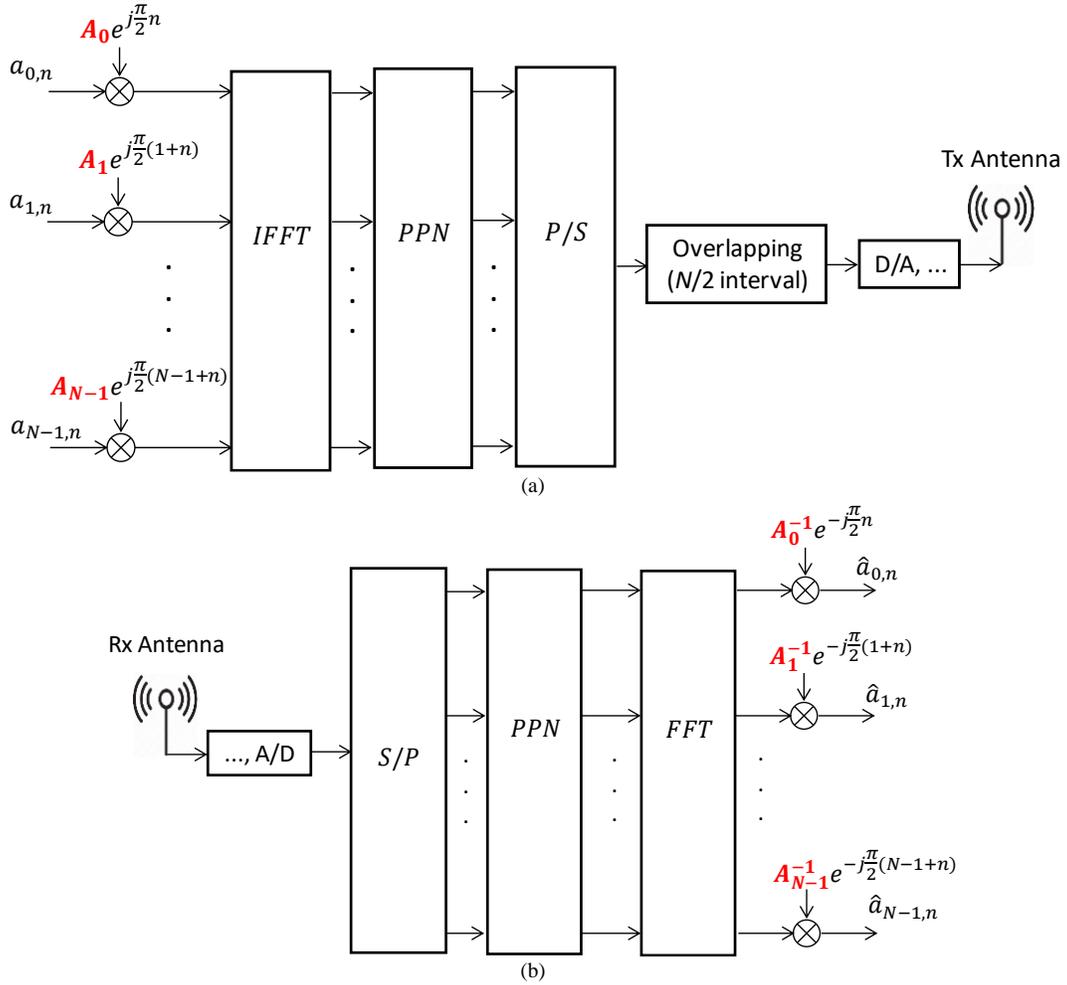

Fig. 3. SS-FBMC Block Diagram, (a) Transmitter (b) Receiver.

The main physical layer characteristics of the proposed SS-FBMC/OFDM and LDACS systems are summarized in Table 1. We note that SS-FBMC parameters are adaptive relying on solution of (3). The total bandwidth of these multicarrier signals is $N\Delta f$, where $N$ is the total number of subcarriers or FFT size and $\Delta f$ is subcarrier spacing.

Table 1. SS-FBMC/OFDM and LDACS physical layer parameters.

|  | SS-FBMC/OFDM | LDACS |
|---|---|---|
| Total bandwidth (MHz) | 1 | 0.625 |
| Effective bandwidth, $B$ (MHz) | varies | ~0.5 |
| # of subcarriers or FFT size ($N$) | $2^l, l =4,5,6,7$ | 64 |
| # of used sub-carriers ($N_u = N-N_g-1$) | varies | 50 |
| # of guard sub-carriers ($N_g$) | varies | 13 |
| Subcarrier spacing $\Delta f$ (kHz) | varies | 9.765 |
| Total symbol duration $T_s$ (µs) | varies | 120 |
| Symbol duration w/o CP (µs) | varies | 102.4 |
| Total guard time $T_g$ due to CP (µs) | SS-FBMC: 0, SS-OFDM: varies | 17.6 |

Parameter $B$ represents the effective RF channel bandwidth. Due to guard subcarriers on both sides of the signal, an effective RF bandwidth of $B = (N_u + 1)\Delta f$ is obtained, where $N_u$ is the number of active or "used" data subcarriers. As mentioned, in SS schemes, following LDACS we assumed the effective bandwidth never exceeds 0.5 MHz, therefore for SS-FBMC/OFDM, we always have $B \leq 0.5$ MHz. We note again that $B$ is not necessarily fixed as in LDACS and could be smaller, depending on the DME power level at the SS-FBMC receiver. Thus, as will be shown in our results, in order to satisfy AWGN BER performance, for high DME powers at SS-FBMC/OFDM receivers, the effective bandwidth might be appreciably smaller.

In Table 1, comparing to LDACS, SS schemes have more flexibility in parameter selection. We chose the entire DME channel bandwidth of 1 MHz instead of the 625 kHz width of LDACS. We have three reasons for this: first, the SS algorithm can adaptively find the guard and data subcarriers based on DME PSD at the receiver; second, it is possible that in some geographic regions, some interferers (e.g., DME channels) may be off, i.e. using different frequency channels; and third, as noted in Table 1, we allow different number of subcarriers ($N = 2^l, l = 4,5,6,7$), where $N$ can be also selected adaptively based on the channel conditions in different environments.

Note that in our current algorithm we do not consider adaptive selection of subcarriers, and all our simulation results are for $N = 128$. In future work we can investigate adaptive selection of the number of subcarriers based on different aeronautical channel scenarios (i.e., using a smaller number of subcarriers during en-route flight phases, because of the more frequency-flat nature of the multipath channel and larger Doppler).

## V. SIMULATION RESULTS

In the following simulations, we assume AWGN and free-space path-loss for link budget calculations. Figure 4 shows the aeronautical communication scenario that we used for simulations. For these simulations, we assume an interferer ground station (GS) such as DME, and two aircraft flying with SS transmitter (Tx) and receiver (Rx).

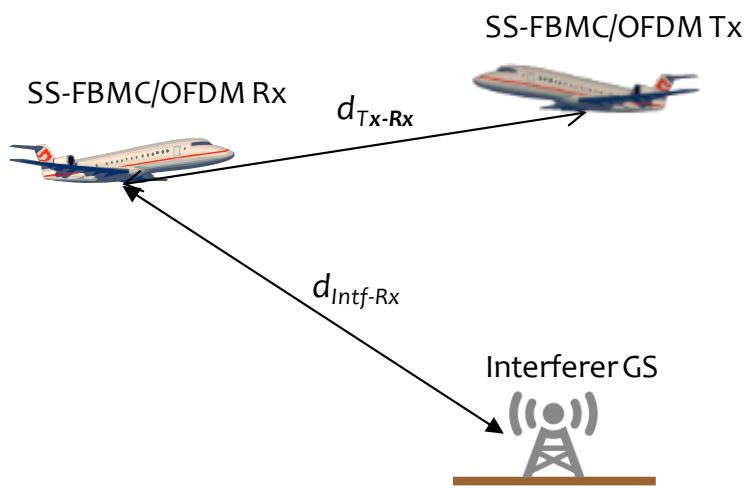

Fig. 4. Aeronautical communication scenario used for simulations.

In simulations we chose the parameters listed in Table I for SS-FBMC/OFDM and LDACS, $P_T = 1$ W, $N = 128$, and 16-QAM modulation. For interference signals we chose the parameters as described in Section III. For SS-FBMC we used the widely used PHYDYAS prototype filter [23], with overlapping factor 4. Also, in these simulations we assumed identical DME or rectangular pulse channels on *both* sides of the SS-FBMC/OFDM signal spectrum with equal power levels.

Figures 5 and 6 show example PSD results for an arbitrary scenario where $d_{Tx\text{-}Rx}$ = 10 km and $d_{Intf\text{-}Rx}$ = 200 km. We chose $K$=1 in (3), as it can satisfy AWGN channel performance for both interference scenarios (refer to Figures 8 and 9). As results show, our algorithm can perfectly find the spectrum locations to transmit subcarriers to satisfy AWGN channel BER performance. For DME, our algorithm found *all* the subcarriers (within the 0.5 MHz allowable bandwidth) as non-interfered subcarriers, but for rectangular pulses, even with the same power as the DME signal, the algorithm could only find 14 subcarriers suitable for data transmission.

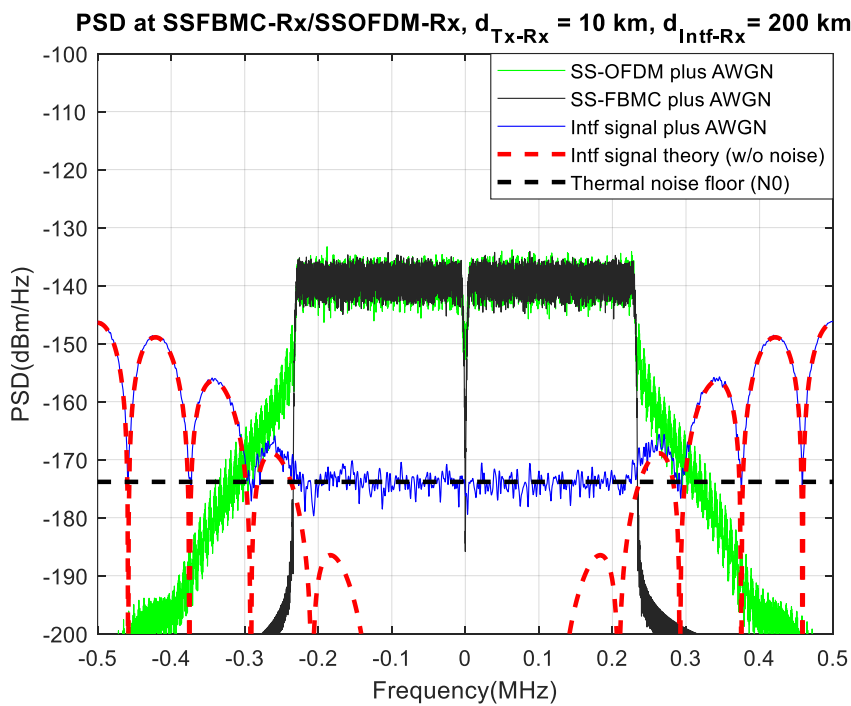

Fig. 5. PSD at SS-FBMC/OFDM Rx after optimization problem solution using $K = 1$, along with *DME Pulses* interference.

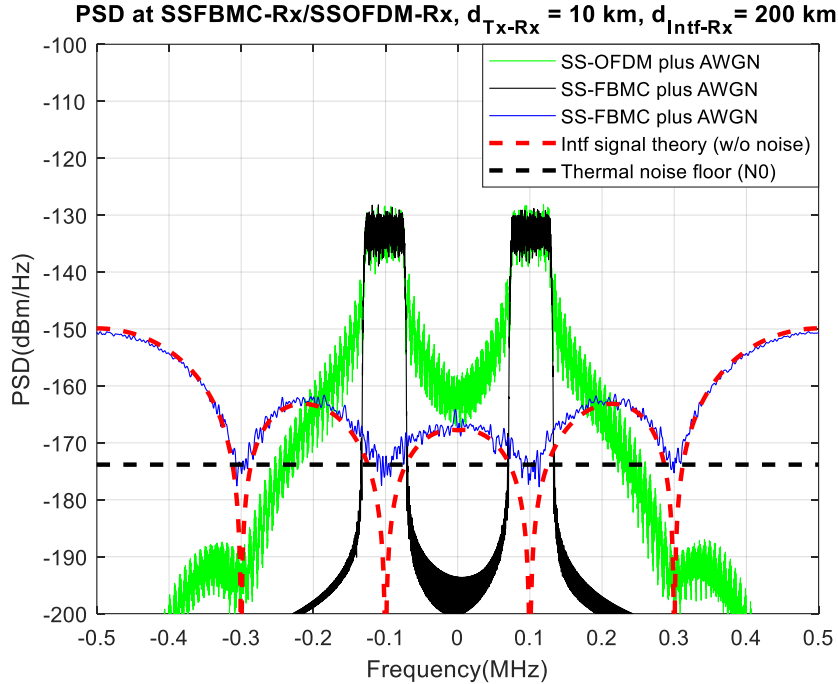

Fig. 6. PSD at SS-FBMC/OFDM Rx after optimization problem solution using $K = 1$, along with *rectangular pulses* interference.

Figure 7 shows the BER performance curves for a scenario where $d_{Tx\text{-}Rx}$ decreases from 180 km to 60 km. This $d_{Tx\text{-}Rx}$ distance range provides $E_b/N_0$ with range of approximately 0-10 dB at SS receivers. Note that for SS schemes we again chose $K = 1$. According to these results, the algorithm satisfies the AWGN channel theoretical performance for both non-white interference signal scenarios and the entire SS scheme flight range. Note that in Figure 7 we only show the SS-FBMC results, and we observed that SS-OFDM provides identical results, but of course has a broader spectrum (see Figures 5 and 6). In Figure 7, for the exact same scenario, we also simulated LDACS with DME interference at two different DME to Rx distances. Results show significant performance degradation compared with the SS schemes, with worst performance at smaller interferer to receiver distances $d_{DME\text{-}Rx}$.

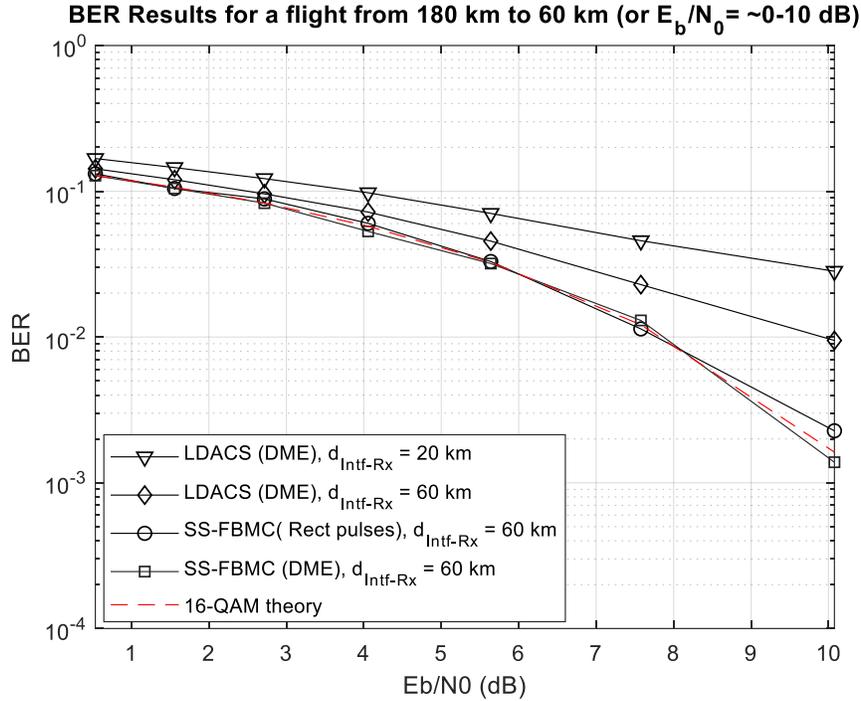

Fig. 7. BER vs. $E_b/N_0$ for $d_{Tx\text{-}Rx}$ = 60 km - 180 km, 16-QAM.

In Figures 8 and 9 we show simulated performance of SS-FBMC (results for SS-OFDM are identical hence not shown) considering different values of $K$ in (3), for DME and rectangular interference scenarios, respectively. Results show that for different interference scenarios, the minimum value of $K$ which attain AWGN performance is different.

For this example, we found $K = 1$ for DME and $K = 3$ for rectangular pulses attaining AWGN performance. We note that via numerous simulations we found $K = 1$ as the minimum value that provides theoretical performance in DME interference.

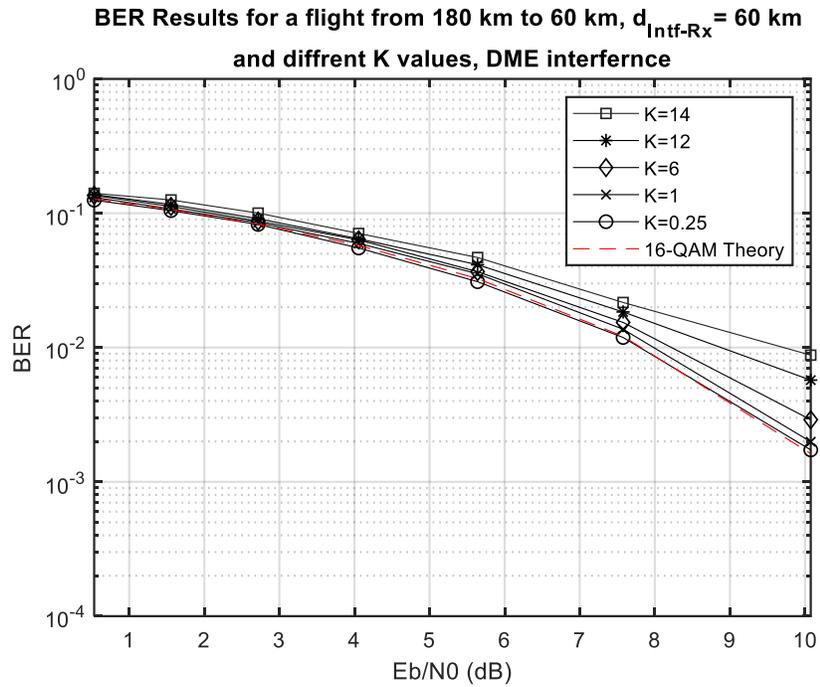

Fig. 8. BER vs. $E_b/N_0$ for different *K* values, and *DME pulses* interference, 16-QAM.

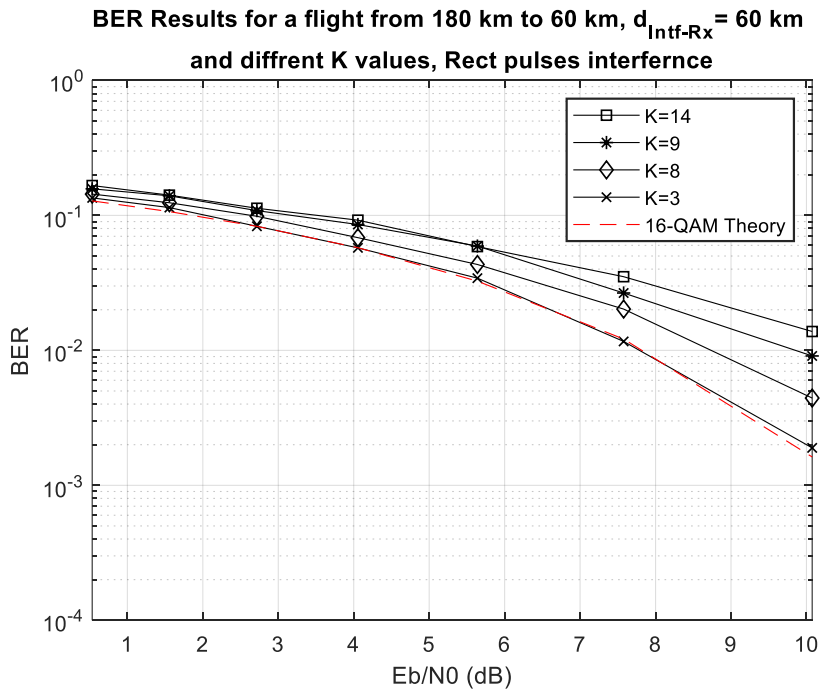

Fig. 9. BER vs. $E_b/N_0$ for different *K* values, and *rectangular pulses* interference, 16-QAM.

To compare the spectral efficiency (bits/sec/Hz) calculated from (3) for different $K$'s, we use the same scenario where $d_{Tx\text{-}Rx} = 60$ km, $d_{Intf\text{-}Rx} = 60$ km. Figure 10 shows the results for DME and rectangular pulse interference. We also include theoretical capacity of LDACS, 6.04 bits/sec/Hz for comparison. Figure 10 shows that our SS algorithm, in addition to attaining AWGN channel BER, can also improve the spectral efficiency, i.e. using $K = 1$ in (3) provides 6.42 bits/sec/Hz, which is approximately a 6 percent increase over LDACS.

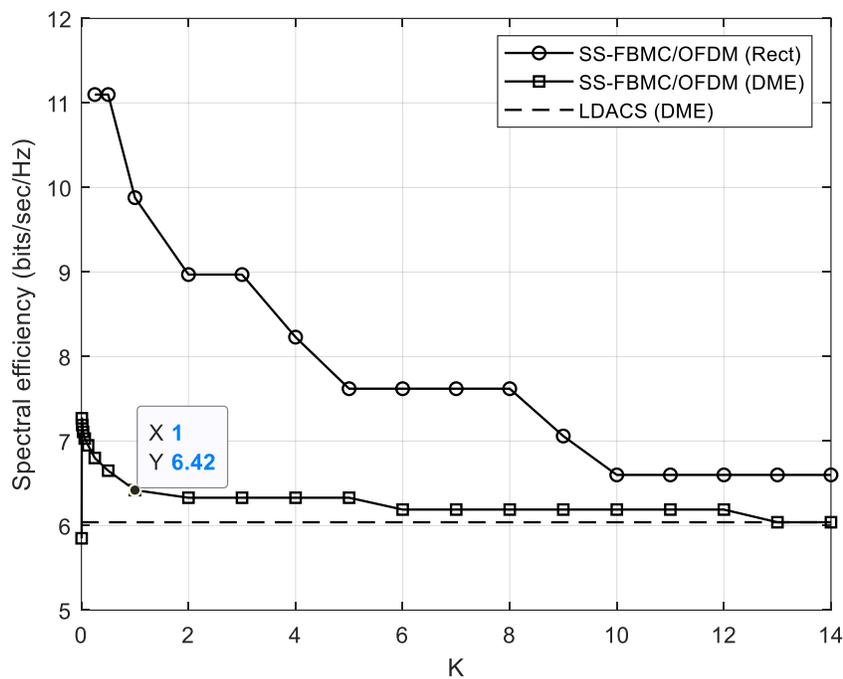

Fig. 10. Capacity versus $K$ for $d_{Tx\text{-}Rx} = 60$ km, $d_{Intf\text{-}Rx} = 60$ km and 16-QAM.

## VI. CONCLUSION

In this paper, we proposed spectrally shaped multicarrier communication schemes using an optimization power allocation algorithm in order to attain theoretical AWGN channel BER performance in the presence of non-white interference or colored-noise channels. For the non-flat spectrum channels, we chose two examples: DME and rectangular pulse signals. According to the proposed algorithm, we solve the optimization problem in order to find the suitable subcarriers for transmitting using the total available power, and we consider the non-suitable subcarriers as null or guard subcarriers. Via simulations, we showed that spectrally shaped schemes can perfectly provide AWGN channel performance in different scenarios, where flat and fixed spectrum schemes such as LDACS incur significant performance degradation. SS approach can also increase the spectral efficiency. Our results show that this algorithm works in non-white interference scenarios, and we show that as an adaptive radio application, it could benefit the future L-band aviation communication systems, enhancing their reliability and capacity. As future work, we can improve the algorithm for an adaptive total number of subcarriers according to the flight and channel condition, including determination of relative ranges where SS is most (and least) effective.

## APPENDIX

Following, we provide the water-filling algorithm proposed to solve (3). We define $C_0$ and $C_1$ variables respectively to check the number of negative and positive assigned power subcarriers.

---

**Algorithm** SS water-filling algorithm

---

1: **Initialization:**

2: **read** $N$, $P_T$, $BW$, $P_{Peak}$, $S(f)$, $N_0 = \sigma^2_{AWGN}$, $K$

3: define $\Delta f = BW/N$

4: **for all** $n \in \{1, 2, \ldots, N\}$ calculate

$I_n = \int_{f_n - \frac{\Delta f}{2}}^{f_n + \frac{\Delta f}{2}} |S(f)|^2 \, df,$

$C2IN_n = \frac{1}{I_n + \Delta f N_0},$

$P_n^{init} = \frac{P_T + \sum C2IN}{N} - C2IN_n.$

5: define $C_0 = \sum n | P_n^{init} < 0$, $C_1 = \sum n | P_n^{init} > 0$

6: **Main loop**:

7: **while** $C_0 \neq 0$ or $C_1 = N$ **do**

8:     **for all** $n \in \{1,2,\ldots,N\}$ define $P_n^{update} = P_n^{init}$

9:     set $P_n^{update} = 0$ **for all** $n$ **if** $P_n^{update} < 0$

10:    set $P_n^{update} = 0$ $\forall n \in \{1, 2, \ldots, \frac{N}{4}+1, \frac{N}{2}+1, \frac{3N}{4}+1, \ldots, N\}$

11:    set $P_n^{update} = 0$ **for all** $n$ **if** $I_n > K\Delta f N_0$

12:    define $N_{rem} = \sum n \mid P_n^{update} \neq 0$

13:    **for all** $m \in \{1,2,\ldots,N_{rem}\}$, and **for all** $n$ **if** $P_n^{update} \neq 0$ define $C2IN_m^{update} = C2IN_n$

14:    define $P_m^{temp} = \frac{P_T + \sum C2IN_m^{update}}{N_{rem}} - C2IN_m^{update}$

15:    **for all** $n$ **if** $P_n^{update} \neq 0$ set $P_n^{update} = P_m^{temp}$

16:    update $C_0 = \sum n \mid P_n^{update} < 0$, $C_1 = \sum n \mid P_n^{update} > 0$

17: **end while**

18: save $P_n^{update}$

Next, we also provide the CVX MATLAB routine to test the Algorithm. As noted, this CVX program, which is designed to solve the convex optimization problems, gives the same results as our proposed algorithm.

**CVX MATLAB** routine for SS water-filling algorithm

C2IN = 1/ (N0*deltaf +I);

Null_subs1= find (I > K * deltaf *N0);

Null_subs2 = [1: N/4+1, N/2+1, 3*N/4+1: N];

Null_subs = union (Null_subs1, Null_subs2);

cvx_begin

```
    variable Pn (1, N)

        Shannon_Cap = deltaf *sum_log (1+Pn.* C2IN);

        objective = Shannon_Cap;

        maximize (objective)

            subject to

                Pn >= 0

                Pn.*C2IN >= 0

                Pn (Null_subs).*C2IN (Null_subs) <= 1e-20

                sum (Pn) == total_Power

cvx_end
```

## ACKNOWLEDGEMENT

This work was supported in part by NASA, under award NNX17AJ94A. The authors also acknowledge Dr. Cenk Erturk for his contributions regarding link budget calculations.